\documentclass[11pt]{article}

\usepackage[preprint]{acl}

\usepackage{times}
\usepackage{latexsym}
\usepackage{listings}
\usepackage{algorithm}
\usepackage{tabularx}
\usepackage{graphicx}
\usepackage{algorithm}
\usepackage{algorithmic}
\usepackage{amsmath}
\newcommand{\TRY}{\item[\textbf{try}]}
\newcommand{\CATCH}[1]{\item[\textbf{catch} (#1)]}

\usepackage{float}

\definecolor{keycolor}{RGB}{31, 119, 180}    
\definecolor{stringcolor}{RGB}{44, 160, 44} 
\definecolor{numcolor}{RGB}{214, 39, 40}    

\usepackage[T1]{fontenc}

\usepackage[utf8]{inputenc}

\usepackage{microtype}

\usepackage{inconsolata}

\usepackage{graphicx}
\graphicspath{{./}{latex/}}

\title{DynAMO: Dynamic Asset Management Orchestration via Topological Multi-Agent Scheduling}

\author{
  Kanishk Kushwaha\textsuperscript{1} \quad
  Harsh Vardhan\textsuperscript{1} \quad
  Dhaval C. Patel\textsuperscript{2} \quad
  Vikrant Vinod Bansode\textsuperscript{1} \\
  \textsuperscript{1}\,Gati Shakti Vishwavidyalaya \quad
  \textsuperscript{2}\,IBM Research \\
  \small{\textbf{Correspondence:} \href{mailto:kanishkkushwaha1234@gmail.com}{kanishkkushwaha1234@gmail.com}}
}

\usepackage{graphicx}
\usepackage{comment}
\usepackage{float} 
\begin{document}
\maketitle
\begin{abstract}

While LLM-powered agents offer end-to-end automation for industrial asset lifecycles, real-world Industry 4.0 deployment is hindered by latency, concurrency instability, and safety risks. We present \textbf{DynAMO} (Dynamic Asset Management Orchestration), a deployment-ready engine utilizing a Plan-then-Execute architecture to generate verifiable workflow graphs. DynAMO supports both \textit{SequentialWorkflow} (topological execution) and \textit{ParallelWorkflow} (dependency-aware concurrency). By dynamically identifying independent tasks, DynAMO preserves structural correctness and safety while significantly improving efficiency through controlled reasoning overlap.

Across six controlled experiments, DynAMO demonstrates substantial performance and robustness gains. Parallel execution reduces end-to-end latency by a median of \textbf{1.6$\times$} over sequential orchestration, rising to \textbf{1.8$\times$} on highly parallelizable workflows. After instrumenting external tool calls with realistic latencies, decomposition shows that LLM reasoning and orchestration still account for more than \textbf{90\%} of execution time, identifying model inference as the primary system bottleneck. Structured context pruning reduces inference latency by approximately \textbf{30\%}, and DynAMO maintains correct functional behaviour (task completion, agent sequencing, and output quality) while exhibiting graceful degradation under controlled fault injection. Reproducibility analysis further confirms stable execution under repeated runs, with parallel scheduling reducing latency variance. These findings establish DynAMO as a practical blueprint for scalable, safe, and latency-aware agent deployment in Industry 4.0 automation pipelines. Code is available at: \url{https://github.com/kushwaha001/DynAMO}.
\end{abstract}

\section{Introduction}
The transition toward Industrial AI demands autonomous systems capable of managing the industrial asset lifecycle through multi-step reasoning and tool-augmented interaction. Industrial agentic tasks require sustained interaction with external environments, iterative information gathering under partial observability, and adaptive strategy refinement. While Large Language Models (LLMs) have shown promise in agentic reasoning frameworks such as ReAct \citep{yao2023reactsynergizingreasoningacting}, deploying them in safety-critical industrial infrastructure requires stronger guarantees of correctness, latency control, and verifiability.

Industrial evaluation presents two primary challenges: (i) agents must operate within complex tool-augmented environments rather than pure text settings, and (ii) workflows produce unstructured outputs such as diagnostic reports and work-order plans, making automated verification non-trivial. To ground our study, we use \textbf{AssetOpsBench} \citep{agentic2025frameworks}, a benchmark providing 141 industrial queries within a simulated IoT-backed environment integrating specialized agents for data retrieval (IoT), failure-mode mapping (FMSR), time-series analysis (TSFM), and work-order generation (WO). Each query (or \emph{utterance}) is a natural-language industrial request that must be decomposed into a multi-step tool-using workflow; for example, ``\textit{Investigate abnormal vibration in Chiller~A and recommend a maintenance action},'' which requires retrieving sensor telemetry, detecting anomalies, mapping them to failure modes, and drafting a work order. This widely recognized benchmark offers realistic industrial scenarios, encompassing the full life-cycle management of critical assets such as wind farms and data centers.

Within this setting, we introduce \textbf{DynAMO} (Dynamic Asset Management Orchestration), a deployment-oriented workflow engine based on a Plan-then-Execute architecture. DynAMO enforces schema-constrained planning to generate verifiable task graphs before execution and employs a topological execution engine that dynamically parallelizes independent tasks. This design improves latency while preserving structural correctness and safety \cite{lu2017industry}. We evaluate DynAMO through six controlled experiments measuring latency reduction, concurrency scalability, context efficiency, fault tolerance, and reproducibility. Our empirical study shows that DynAMO achieves a \textbf{median 1.6$\times$ latency reduction} (up to 1.8$\times$ on highly parallelizable workflows; see Table~\ref{tab:exp1_latency}) via parallel execution, identifies LLM reasoning and orchestration as the dominant cost component (>90\% after tool-I/O instrumentation; Section~\ref{subsec:exp2}), quantifies scalability limits under concurrency stress, and shows robustness under controlled fault injection.

Our contributions are threefold:
\begin{itemize}
    \item \textbf{System design.} We present a deployment-ready Plan-then-Execute engine that combines schema-constrained planning with tool-aware, dependency-driven topological scheduling. Unlike orchestration libraries that treat planning and execution as loosely coupled steps, DynAMO validates the full task graph at planning time and uses agent–tool metadata to overlap I/O-bound and compute-bound work without violating dependencies.
    \item \textbf{Empirical characterization.} On the AssetOpsBench industrial benchmark, we provide a fine-grained, real-execution study—latency decomposition, concurrency stress testing, context-size sensitivity, fault injection, and run-to-run reproducibility—that is more thorough than typical agent-system evaluations, and we report functional-correctness metrics alongside latency.
    \item \textbf{Deployment insight.} We show that, once external tool latency is properly accounted for, LLM inference and orchestration still dominate end-to-end cost (>90\%). This reframes the engineering priority for industrial agent systems from orchestration micro-optimization toward inference- and context-efficiency, and motivates benchmarking deployed reliability rather than leaderboard accuracy alone.
\end{itemize}

We position DynAMO not as a new individual technique but as a benchmark-grounded study of what it takes to make schema-constrained planning, controlled parallelism, and context-efficient execution reliable under real industrial constraints, offering concrete guidelines for benchmarking and real-world Industry~4.0 deployment.

\section{Related Work}

\paragraph{LLM-based Agentic Frameworks-}
Recent work has explored autonomous LLM agents capable of tool use and multi-step reasoning, including ReAct-style prompting \citep{yao2023reactsynergizingreasoningacting} and multi-agent orchestration systems \citep{Anthropic2024Agents}. These frameworks demonstrate strong reasoning capabilities but typically assume loosely structured workflows and do not enforce explicit dependency validation or safety constraints required in industrial environments.

Declarative orchestration approaches have introduced hybrid sequential–parallel execution patterns \citep{daunis2025declarativelanguagebuildingorchestrating}, improving flexibility compared to static pipelines. However, they generally lack formal mechanisms for verifying dependency correctness or preventing malformed execution graphs prior to runtime.

\paragraph{Benchmarking Agent Reliability-}
In the emerging era of benchmarks, several industry initiatives are helping to accelerate research innovation by building suites such as AssetOpsBench \cite{agentic2025frameworks}, MCP-Universe \cite{luo2025mcpuniversebenchmarkinglargelanguage}, and others. However, these benchmarks mainly focus on agentic capability rather than the latency, concurrency, and reliability concerns that govern product-level deployment. Our work instead asks how such benchmarks can be leveraged to study latency and real-time deployment behaviour.

Recent evaluations of agentic systems reveal that increasing architectural complexity often leads to degraded tool selection accuracy, redundant reasoning loops, and inflated latency \citep{benchmarks2025emerging}. These findings highlight the fragility of multi-step LLM workflows, particularly under extended context and nested dependencies.

\paragraph{System Efficiency and Practical Constraints-}
Practical analyses of orchestration frameworks such as LangChain, LangGraph, and CrewAI report significant variability in token overhead and execution latency \citep{mazzolenis2025agentwarppworkflowadherence}. Heavy memory tracking and sequential execution can introduce substantial inefficiencies,\cite{wang2022self} limiting deployment in latency-sensitive environments.

\paragraph{Gap and Positioning-}
Existing systems either emphasize flexible orchestration without structural guarantees or focus on benchmarking without addressing deployment-level safety. General-purpose frameworks such as LangGraph and CrewAI do expose parallel execution modes, but they leave dependency correctness, plan validation, and context propagation largely to the developer. Our goal is not to introduce a fundamentally new scheduling primitive, but to integrate (i) schema-validated planning, (ii) explicit DAG-based dependency verification, (iii) dynamic, tool-aware parallel execution, and (iv) runtime fault isolation into a single engine, and to characterize its behaviour under realistic industrial load. DynAMO thus targets the deployment gap: making known orchestration ideas verifiable, reproducible, and latency-characterized for Industry~4.0 settings. We discuss the absence of head-to-head comparison against these frameworks as an explicit limitation (Section~\ref{sec:limitations}).
\section{Core Components of DynAMO}

To operationalize this vision, DynAMO introduces three architectural innovations designed specifically for the constraints of industrial deployment:

\subsection{Schema-Constrained Planning}
\label{subsec:planner_mechanism}

DynAMO enforces structural correctness at planning time using a
\textbf{schema-constrained planner}. Instead of generating free-form text,\cite{huang2025survey}
the LLM must output a workflow that conforms to a predefined JSON schema (See Appendix \ref{app:sche}). Figure \ref{fig:track1} gives a high level summary. 

\paragraph{Schema Format-} Each plan must follow:

\begin{lstlisting}[caption={Task dependency JSON specification.}]
{
  "tasks": [
    {"id": "S1", "desc":"...", "agent": "IoT",  "deps": [], "out":"..."},
    {"id": "S2", "desc":"...", "agent": "TSFM", "deps": ["S1"], "out":"..."},
  ]
}
\end{lstlisting}

\begin{figure*}[t]
    \centering
    \begin{minipage}{0.28\textwidth}
        \centering
        \includegraphics[width=\linewidth]{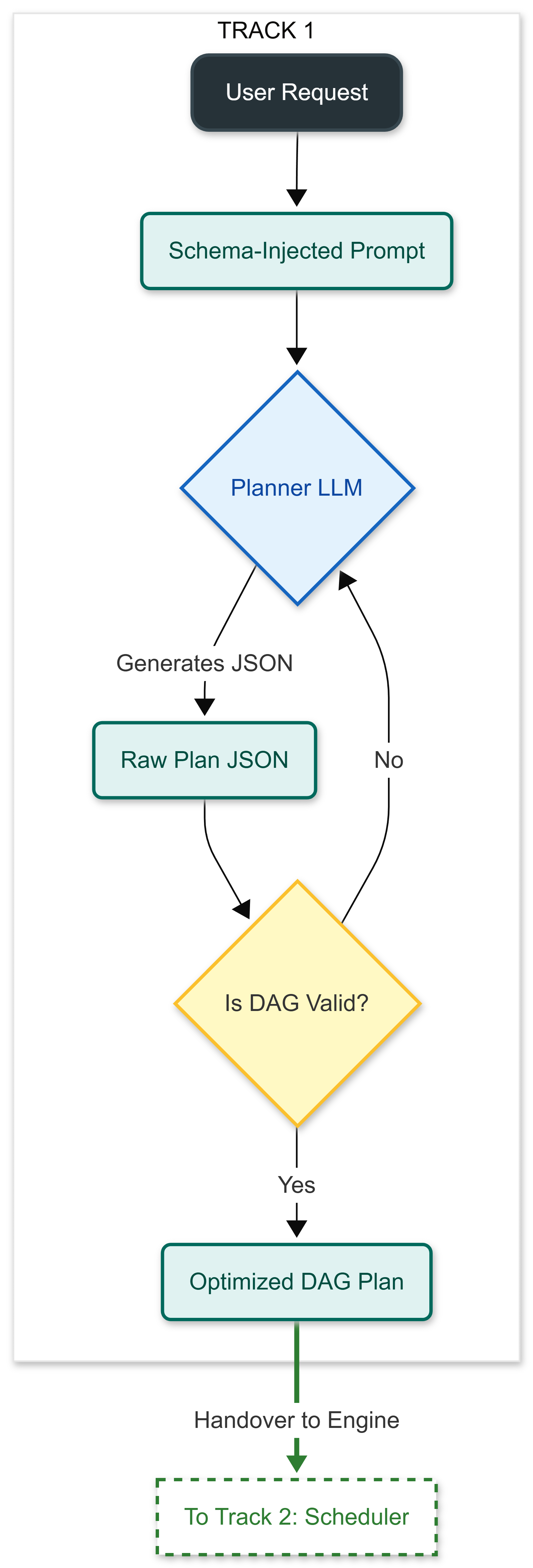}
        \caption{ Schema-Constrained Planning}
        \label{fig:track1}
    \end{minipage}
    \hfill 
    \begin{minipage}{0.68\textwidth}
        \centering
        \includegraphics[width=265.05bp]{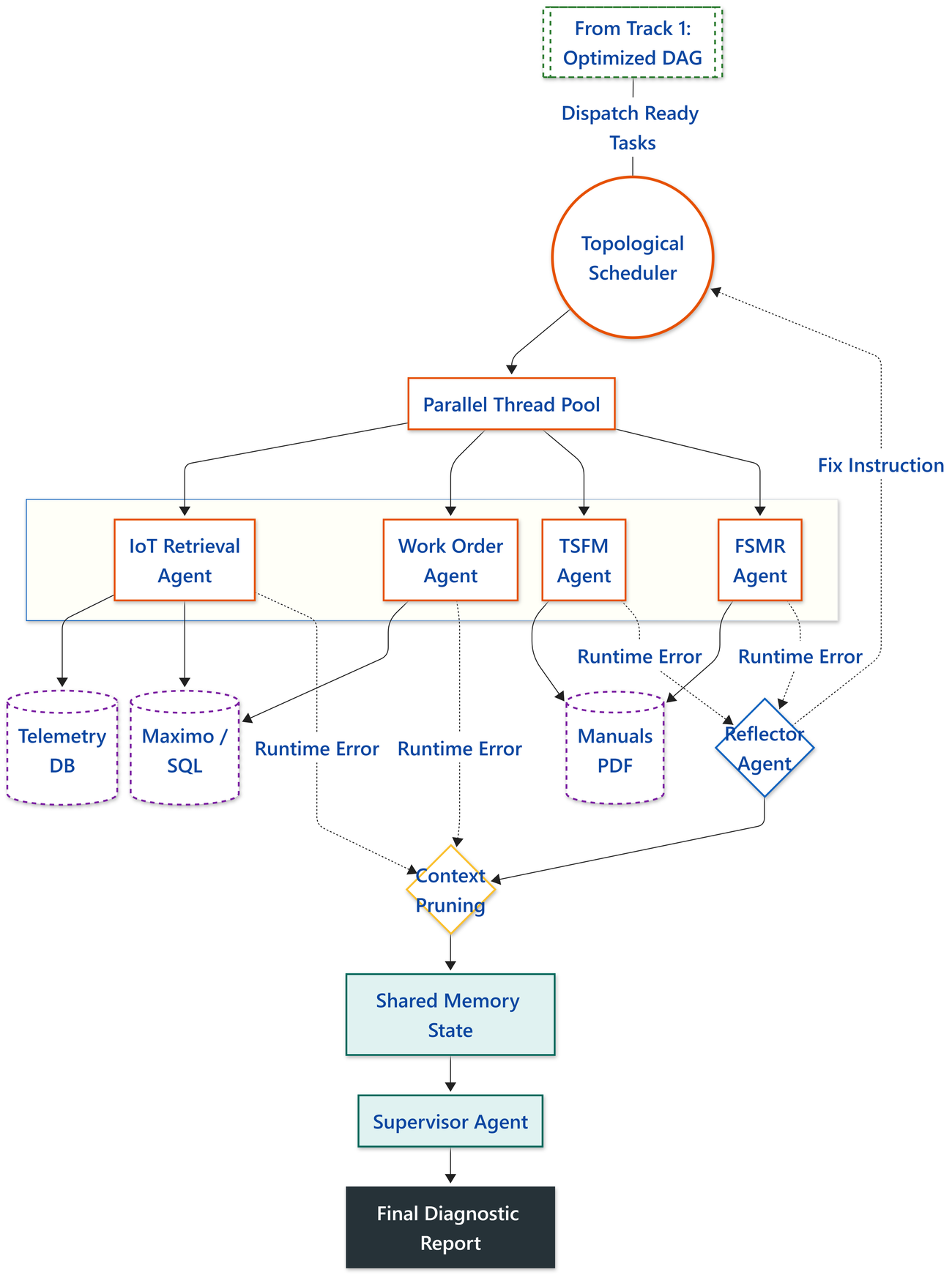}
        \caption{ Topological Execution Engine: Building Block consists of Agents, Dataset and Thread Pool}
        \label{fig:track2}
    \end{minipage}
\end{figure*}

The schema enforces: (i) valid task identifiers,
(ii) registered agent names,
(iii) explicit dependencies,
(iv) declared outputs.
Concretely, the planner's raw JSON output is parsed and checked against this schema: each \texttt{agent} field must resolve against a fixed \emph{agent registry} (IoT, TSFM, FMSR, WO), each entry in \texttt{deps} must reference a previously declared task id, and the induced dependency graph must be acyclic. If any check fails, the plan is rejected and the planner is re-prompted with the specific violation, up to a bounded retry budget (5 in our configuration); persistent violations fall back to safe defaults (Appendix~\ref{app:sche}). This converts plan validity from a runtime hope into a pre-execution guarantee.

\paragraph{Execution Graph-}
The validated plan is converted into a directed acyclic graph (DAG)\ref{fig:track1},
where tasks are nodes and \texttt{deps} define edges\cite{leong2025dynaswarmdynamicallygraphstructure}.
Independent nodes execute in parallel; dependent nodes wait.

\paragraph{Example 1-}
For the query:

\emph{``Investigate abnormal vibration in Chiller A.''}

\begin{lstlisting}[caption={Task dependency for Example 1.}]
{
  "tasks": [
    {"id": "S1", "agent": "IoT",  "deps": []},
    {"id": "S2", "agent": "TSFM", "deps": ["S1"]},
    {"id": "S3", "agent": "FMSR", "deps": []},
    {"id": "S4", "agent": "FMSR", "deps": ["S2", "S3"]}
  ]
}
\end{lstlisting}

Here, S2 and S3 execute concurrently, while S4 waits for both.
This structure enables parallelism without compromising correctness.

\paragraph{Safety Guarantee-}
By validating structure before execution,\cite{madaan2023self,wang2022self} DynAMO prevents
invalid agent assignments, cyclic dependencies, and undefined references,
shifting error detection from runtime to planning time.


\subsection{Topological Execution Engine}
\label{subsec:execution_engine}

Industrial diagnostics frequently require retrieving data from multiple
independent assets (e.g., querying telemetry for one chiller while
analyzing failure logs for another). Sequential execution introduces
artificial bottlenecks, leaving CPU resources idle during I/O waits.

DynAMO employs a \textbf{topological execution engine} that transforms the
validated DAG (See Appendix \ref{app:top}) into a dynamically scheduled dependency graph. The engine:

\begin{enumerate}
    \item Computes in-degree dependencies to identify \emph{ready} tasks.
    \item Classifies tasks based on their assigned tools
          (e.g., I/O-bound database queries vs.\ CPU-bound model inference).
    \item Dispatches tasks to a parallel thread pool with resource-aware scheduling.
\end{enumerate}

Unlike naive parallelism, the scheduler leverages agent–tool metadata to
differentiate execution characteristics. For example, telemetry retrieval
(I/O-bound) can be overlapped with anomaly detection (CPU-bound model
inference), maximizing throughput while minimizing idle cycles.  Figure \ref{fig:track2} gives brief architecture. 

If a non-critical branch encounters a runtime failure (e.g., API timeout),
the engine isolates the exception and continues executing unaffected
branches. This ensures fault containment without cascading termination.

By combining dependency-aware topological ordering with tool-aware
resource scheduling, DynAMO converts linear agent pipelines into
concurrent, fault-tolerant execution graphs optimized for Industry~4.0
constraints.

\paragraph{Token-Budgeted Context Pruning:}
A major barrier to scaling multi-agent systems is ``context saturation,'' where voluminous industrial logs exceed the effective window of LLMs, often leading to the ``lost-in-the-middle'' phenomenon in which critical details placed away from the prompt boundaries are under-used \cite{liu2023lost}. We address this via a \textbf{Token-Budgeted Pruning} mechanism that enforces a \textbf{dynamic token cap} at each agent handover. Instead of blindly propagating raw telemetry blobs or large SQL dumps, DynAMO passes \textbf{lightweight schema headers} and \textbf{immutable file pointers} between agents. When reasoning agents require specific data points, the system employs \textbf{just-in-time retrieval} coupled with a \textbf{semantic relevance scorer} to fetch only the most pertinent snippets based on vector similarity (Appendix~\ref{app:context}). This keeps the model's attention focused on high-signal content rather than diluting it with irrelevant context. Empirically, this structured context passing lowers reasoning latency by approximately \textbf{30\%} relative to propagating raw telemetry (Section~\ref{sec:exp_context_latency}, Table~\ref{tab:exp4_context}), reducing inference cost while preserving the contextual fidelity required for multi-step root-cause analysis (fault handling detailed in Appendix~\ref{app:faulty}).




\section{Experiments}
\label{sec:experiments}

This section evaluates the performance, scalability, and robustness of agentic workflows under different execution and operating conditions. We conduct six experiments that progressively analyze execution strategy, latency composition, concurrency limits, context efficiency, and fault tolerance.

\paragraph{Experimental Setup-}
\label{subsec:setup}

All experiments are conducted in a real execution environment using production-grade agents and tools, over the \textbf{141 industrial queries} of AssetOpsBench. Each workflow is generated by an LLM-based planner and then executed. The planner and all reasoning agents use \texttt{meta-llama/\allowbreak Llama-4-Maverick-\allowbreak 17B-128E-Instruct-FP8} at temperature $0.0$ (deterministic planning), with at most 5 plan-regeneration retries and 15 workflow steps (full configuration in Appendix). To reduce stochastic noise, all reported values correspond to the \textbf{median over three independent runs}.

Latency is measured using wall-clock time and decomposed into the following components:

\begin{itemize}
    \item \textbf{Planner latency}: Time taken by the LLM planner to generate the task graph.
    \item \textbf{Reasoning latency}: Time spent in LLM-driven agent reasoning.
    \item \textbf{Tool I/O latency}: Time spent executing external tools.
    \item \textbf{Synchronization overhead}: Residual execution time not attributed to the above components.
\end{itemize}

\paragraph{Workflow Graph Complexity.}

Across the 141 evaluated industrial queries, the generated workflow graphs were not limited to linear chains. 
On average, each plan contained 3--5 tasks, with approximately 40\% of workflows exhibiting at least one parallelizable branch (i.e., multiple nodes with zero in-degree after dependency resolution). 

Typical structures included:
(i) independent data retrieval and failure-mode lookup tasks executed concurrently, and 
(ii) anomaly analysis and mapping steps merged via a final aggregation node.

Thus, DynAMO did not operate on trivial single-path DAGs. Instead, a substantial fraction of plans contained multi-branch dependency structures that enabled measurable parallel execution gains.

\subsection{Experiment 1: Sequential vs.\ Parallel Execution}
\label{subsec:exp1}
This experiment compares end-to-end workflow latency under sequential and parallel orchestration strategies. Sequential execution enforces strict task ordering, whereas parallel execution allows independent tasks to execute concurrently subject to dependency constraints.
\paragraph{Results-}
Parallel execution consistently outperforms sequential execution. On the aggregate latency reported in Table~\ref{tab:exp1_latency}, the median speedup is \textbf{1.62$\times$} ($308.53\text{s} \rightarrow 190.49\text{s}$). The gain is not uniform across queries: workflows that are essentially linear chains see little improvement, whereas highly parallelizable workflows (multiple zero in-degree branches) reach up to \textbf{1.8$\times$}. We therefore report 1.6$\times$ as the representative (median) figure and 1.8$\times$ as the best-case rather than the headline number. This improvement arises purely from overlapping agent reasoning steps that would otherwise be serialized.

\begin{table}[h!]
\centering
\begin{tabular}{lc}
\hline
\textbf{Execution Mode} & \textbf{Median Latency (s)} \\
\hline
Sequential & 308.53 \\
Parallel & 190.49 \\
\hline
\end{tabular}
\caption{End-to-end latency comparison between sequential and parallel execution.}
\label{tab:exp1_latency}
\end{table}

\paragraph{Key Observation- }
\textbf{Parallel orchestration improves performance without modifying agent logic or planner behavior}, demonstrating that execution strategy alone can yield substantial latency reductions in agentic workflows.

\subsection{Experiment 2: Latency Decomposition}
\label{subsec:exp2}

This experiment analyzes where execution time is spent by decomposing workflow latency into planning, reasoning, tool I/O, and synchronization components.

\paragraph{Instrumentation note-}
In our initial measurements, tool I/O latency was reported as zero because external tool execution was not instrumented for timing (to preserve strict tool schema validation and avoid modifying production tool interfaces). As reviewers correctly noted, this is unrealistic for industrial settings, where database queries and API calls are non-trivial latency sources, and it inflates the apparent share of LLM reasoning. We therefore re-ran the decomposition with external tool calls instrumented using realistic per-call delays (\textbf{0.8--2.5\,s} per tool call, drawn from observed ranges for sensor and history retrieval). We report the corrected decomposition below.

\paragraph{Results-}
After instrumentation, \textbf{LLM inference and orchestration still dominate} end-to-end latency, accounting for roughly \textbf{91\%}, while tool I/O contributes about \textbf{9\%} (Table~\ref{tab:exp2_decomp}). Planner latency is incurred only once per workflow and is negligible by comparison, and synchronization overhead remains minimal and does not increase under parallel execution. The earlier ``>95\%'' estimate was thus a slight overestimate; the corrected figure (>90\%) leaves the qualitative conclusion unchanged.

\begin{table}[h!]
\centering
\begin{tabular}{lc}
\hline
\textbf{Latency Component} & \textbf{Share (\%)} \\
\hline
LLM inference + orchestration & 90.97 \\
Tool I/O (instrumented) & 9.03 \\
\hline
\end{tabular}
\caption{Latency decomposition after instrumenting external tool calls with realistic delays (0.8--2.5\,s per call). LLM inference and orchestration remain the dominant cost.}
\label{tab:exp2_decomp}
\end{table}

\paragraph{Key Observation-}
Even when external tool latency is accounted for realistically, the primary bottleneck in agentic workflows is \textbf{LLM inference and orchestration rather than tool I/O or synchronization}. Parallel execution reduces overall latency by overlapping reasoning phases, not by reducing per-agent reasoning cost. This is the central deployment insight: efforts to speed up industrial agent systems should target inference and context efficiency before orchestration micro-optimization.

\subsection{Experiment 3: Concurrency Stress Testing}
\label{subsec:exp3}

This experiment simulates a realistic deployment scenario in which multiple users issue industrial diagnostic queries concurrently. Unlike earlier experiments that analyze a single workflow instance, this evaluation measures system behavior under simultaneous multi-workflow execution.

\begin{table}[h!]
\centering
\small
\begin{tabularx}{\columnwidth}{>{\centering\arraybackslash}X >{\centering\arraybackslash}X >{\centering\arraybackslash}X >{\centering\arraybackslash}X}
\hline
\textbf{Conc.} & \textbf{Avg.\ Lat.\ (s)} & \textbf{Max Lat.\ (s)} & \textbf{Fails} \\
\hline
1  & 94.75 & 234  & 0.0 \\
5  & 266.0 & 1296 & 0.4 \\
10 & 161.0 & 412  & 0.6 \\
20 & 19.7  & 42   & 19.6 \\
\hline
\end{tabularx}
\caption{Concurrency stress test results. ``Conc.'' is the number of simultaneous workflows; ``Fails'' is the mean number of failed workflows.}
\label{tab:exp3_concurrency}
\end{table}

\paragraph{Failure Analysis.}
Observed failures under high concurrency are primarily attributable to shared LLM inference and external service contention rather than DAG scheduling errors. 
When multiple workflows invoke reasoning or tool calls simultaneously, resource saturation occurs at the model serving layer and API endpoints. This leads to increased latency variance and, at high concurrency levels, request timeouts.

Importantly, the planning and dependency resolution mechanisms remain structurally correct under load. Failures arise from infrastructure-level bottlenecks (e.g., limited LLM throughput and external database I/O limits), not from invalid DAG construction.

\paragraph{Interpretation.}
These results suggest that concurrency-induced degradation is a systems-level phenomenon rather than a logical planning flaw. This finding reinforces the need for admission control, model parallelization, or workload-aware scheduling when deploying LLM-driven agents in Industry 4.0 environments.

\subsection{Experiment 4: Context Size vs.\ Latency}
\label{sec:exp_context_latency}

This experiment evaluates how the size and structure of contextual information affect inference latency.

\paragraph{Setup-}
We evaluate three context representations:\\
\textbf{(i)} Raw Telemetry,\\
\textbf{(ii)} Pruned Summary,\\
\textbf{(iii)} Schema-only Context.\\
Execution is performed in a controlled simulation environment with token-dependent latency.

\begin{table}[H]
\centering
\begin{tabular}{lc}
\hline
\textbf{Context Type} & \textbf{Latency (s)} \\
\hline
Raw Telemetry & 0.585 \\
Pruned Summary & 0.446 \\
Schema-only & 0.412 \\
\hline
\end{tabular}
\caption{Impact of context representation on reasoning latency.}
\label{tab:exp4_context}
\end{table}

\paragraph{Token Statistics.}
Across experiments, the average input context size ranged from approximately 800 tokens (schema-only context) to 12,000 tokens (raw telemetry simulation). For real execution experiments, average prompt size per task was approximately 2,500–3,200 tokens, including system prompts, tool descriptions, and intermediate reasoning context. Output responses averaged 300–800 tokens depending on task complexity.

\paragraph{Results-}
Latency increases monotonically with context size. Schema-only context achieves the lowest latency, while raw telemetry incurs the highest cost.

\paragraph{Key Observation-}
\textbf{Structured context passing reduces inference latency by approximately 30\%} without modifying workflow logic, making it essential for real-time industrial systems.

\subsection{Experiment 5: Fault Injection and Latency Stability}
\label{sec:exp_fault_latency}

This experiment evaluates robustness under partial failures.

\paragraph{Setup-}
We inject two fault types: \\
\textbf{(i)} Tool Timeout, \\
\textbf{(ii)} Partial Sensor Failure.\\
A no-fault execution is used as a baseline.

\paragraph{Results-}
Both faults increase latency, with tool timeouts having the largest impact. However, all workflows complete successfully.

\begin{table}[H]
\centering
\begin{tabular}{lc}
\hline
\textbf{Fault Condition} & \textbf{Latency (s)} \\
\hline
No Fault & 0.445 \\
Partial Sensor Failure & 0.845 \\
Tool Timeout & 1.245 \\
\hline
\end{tabular}
\caption{Latency impact under fault injection.}
\label{tab:exp5_faults}
\end{table}

\paragraph{Key Observation-}
The system exhibits \textbf{graceful degradation}, isolating failures while maintaining bounded latency and successful completion.

\subsection{Experiment 6: Reproducibility and Run-to-Run Stability}
\label{sec:exp_reproducibility}

While average latency provides a coarse measure of performance, real-world industrial deployments require \textbf{stable and reproducible execution behavior}. This experiment evaluates the run-to-run variability of agentic workflows under repeated executions,\cite{paulbellow2025defeating} focusing on latency dispersion rather than raw speedup.

\paragraph{Setup-}
We execute the same workflow multiple times under identical conditions using both \textit{SequentialWorkflow} and \textit{ParallelWorkflow} schedulers. For each configuration, we record the median latency, standard deviation, and extreme values (minimum and maximum latency).

\paragraph{Results-}
While the median speedup of parallel execution is modest (approximately $1.01\times$), parallel execution exhibits \textbf{significantly lower variance}, reducing standard deviation by more than \textbf{2.4$\times$}. Sequential execution shows large performance fluctuations, with latency spanning over 600 seconds between best- and worst-case runs.

\begin{table}[H]
\centering
\setlength{\tabcolsep}{4pt}
\begin{tabular}{lcc}
\hline
\textbf{Metric} & \textbf{Sequential} & \textbf{Parallel} \\
\hline
Median Latency (s) & 802.34 & 793.17 \\
Std. Deviation (s) & 239.06 & 96.28 \\
Minimum Latency (s) & 451.51 & 709.82 \\
Maximum Latency (s) & 1099.33 & 972.38 \\
\hline
\textbf{Median Speedup} & \multicolumn{2}{c}{1.01$\times$} \\
\hline
\end{tabular}
\caption{Reproducibility analysis comparing sequential and parallel execution across repeated runs.}
\label{tab:exp6_reproducibility}
\end{table}

\paragraph{Key Observation-}
\textbf{Parallel orchestration improves execution stability even when average latency gains are small}. Reduced variance and tighter latency bounds are critical properties for industrial agentic systems, where predictability and worst-case guarantees often matter more than marginal improvements in mean performance.

\subsection{Functional Correctness}
\label{sec:exp_accuracy}

Latency and stability are necessary but not sufficient for a deployment-oriented system: a fast workflow that produces wrong plans or wrong answers is not useful. We therefore complement the latency analysis with system-level functional-correctness metrics. On a curated set of \textbf{11 representative multi-step diagnostic workflows} (covering sensor retrieval, anomaly analysis, failure-mode mapping, and work-order generation), we manually scored each execution along four axes, plus a qualitative check for hallucinated tool calls or fabricated values.

\begin{table}[h!]
\centering
\begin{tabular}{lc}
\hline
\textbf{Correctness Dimension} & \textbf{Score} \\
\hline
Task completion          & 8/11 \\
Data retrieval accuracy  & 7/11 \\
Correct agent sequencing & 9/11 \\
Output clarity           & 10/11 \\
\hline
\end{tabular}
\caption{System-level functional correctness on 11 representative multi-step workflows. Schema-constrained planning yields high correct-sequencing and output-clarity rates; data-retrieval errors are the main remaining failure source.}
\label{tab:exp_accuracy}
\end{table}

\paragraph{Analysis-}
Schema-constrained planning produces correct agent sequencing in 9/11 cases and clear, well-formed outputs in 10/11, indicating that plan-time validation effectively prevents malformed or mis-ordered workflows. The lower data-retrieval score (7/11) is the dominant remaining error source and stems largely from upstream tool/query grounding rather than from scheduling, consistent with our latency finding that tool interaction is a real (if not dominant) cost. We also observed reduced hallucinated tool invocations relative to free-form prompting, which we attribute to the explicit agent registry and dependency declarations.

\paragraph{Caveat-}
These scores are computed on a small curated subset and are intended to characterize qualitative failure modes, not to claim state-of-the-art accuracy. The absolute numbers are modest, and full-benchmark accuracy evaluation across all 141 queries (with automated scoring) is left to future work (Section~\ref{sec:limitations}).

\section{Conclusion}

We presented \textbf{DynAMO}, a deployment-oriented orchestration engine for industrial agentic systems that integrates schema-constrained planning with topological, dependency-aware execution. This design enables verifiable workflow construction and safe parallel execution in complex, safety-critical environments.

Experimental results on AssetOpsBench demonstrate that DAG-based parallel execution achieves consistent latency reductions (median 1.6$\times$) without modifying agent logic, while maintaining low orchestration overhead. After instrumenting external tool calls, we find that LLM inference and orchestration still account for more than 90\% of end-to-end latency, reframing the optimization target for industrial agent systems. We further show that structured context passing reduces inference latency, that schema-constrained planning yields high correct-sequencing and output-clarity rates, and that fault isolation enables graceful degradation under partial failures. Concurrency stress testing highlights scalability limits, emphasizing the need for admission control and resource-aware scheduling in real deployments. We do not yet claim superiority over external orchestration frameworks—establishing that comparison is the key next step—but DynAMO shows that combining planning-time verification with execution-time parallelism provides a practical, measurable foundation for reliable Industry~4.0 automation.

\section*{Limitations}
\label{sec:limitations}
While DynAMO improves latency and structural robustness, several limitations remain:
\begin{itemize}
    \item \textbf{No external-framework baseline.} Our latency comparison is between DynAMO's own sequential and parallel schedulers on the same validated task graph. We do not provide head-to-head comparison against the parallel execution modes of general-purpose frameworks such as LangGraph, CrewAI, or AutoGen. Such a comparison is the most important direction for future work and is required to substantiate claims of advantage over existing orchestration systems; the present results should be read as an internal characterization of plan-time validation and dependency-aware scheduling, not as a demonstration of superiority over those frameworks.
    \item \textbf{Limited accuracy evaluation.} Functional-correctness metrics (Section~\ref{sec:exp_accuracy}) are reported on a small curated subset of workflows and are scored manually. Automated, full-benchmark accuracy and fault-diagnosis evaluation across all 141 queries remains future work.
    \item \textbf{LLM dependence.} System performance remains bounded by LLM inference latency and availability, which dominate overall execution time even after accounting for tool I/O.
    \item \textbf{Concurrency constraints.} Under high concurrent load, shared inference resources lead to instability, indicating the need for admission control and stronger isolation mechanisms.
    \item \textbf{Simulation environment.} Experiments are conducted within the AssetOpsBench industrial simulation; real-world deployments may introduce additional operational variability. Tool I/O latencies in Section~\ref{subsec:exp2} are realistic simulated delays rather than measurements from a specific production backend.
\end{itemize}

\section*{Ethical Considerations}
The authors are not aware of any ethical concerns arising from this research. All experiments were conducted on the publicly available AssetOpsBench benchmark within a simulated industrial environment, and no human-subject or personally identifiable data was used.

\bibliography{custom}

\appendix

\newpage
\section{Appendix: Implementation Details}

This appendix provides detailed pseudocode and configuration details omitted from the main paper due to space constraints.


\subsection{ Schema-Constrained Planning}
\label{app:sche}
DynAMO enforces structural correctness at planning time by validating LLM-generated plans against a predefined task schema before execution.

\begin{algorithm}[H]
\small
\caption{Schema-Constrained Planning}
\begin{algorithmic}[1]
\REQUIRE User query $q$, registered agents $\mathcal{A}$
\ENSURE Valid task list $\mathcal{T}$

\STATE $P \gets \text{LLM}(q)$
\STATE Extract tasks, agents, and dependencies from $P$

\IF{schema violation}
    \STATE Regenerate $P$
\ENDIF

\FORALL{tasks $t$}
    \IF{$t.\text{agent} \notin \mathcal{A}$}
        \STATE Assign default agent
    \ENDIF
    \STATE Remove invalid dependencies
\ENDFOR

\IF{cycle detected}
    \STATE Reject and regenerate
\ENDIF

\RETURN $\mathcal{T}$
\end{algorithmic}
\end{algorithm}


\subsection{ Topological Execution Engine}
\label{app:top}
Execution is dependency-aware and parallelized using DAG traversal.

\begin{algorithm}[H]
\small
\caption{Dependency-Aware Parallel Execution}
\begin{algorithmic}[1]

\REQUIRE Directed Acyclic Graph $G=(V,E)$
\ENSURE Execution history $\mathcal{H}$

\STATE $\mathcal{H} \gets \emptyset$
\STATE Compute $\text{in-degree}$ for all nodes $v \in V$
\STATE $R \gets \{v \in V \mid \text{in-degree}(v) = 0\}$ \COMMENT{Initialize ready queue}

\WHILE{$R \neq \emptyset$}
    \STATE $\text{ParallelExecute}(R)$ 
    \STATE $\mathcal{H} \gets \mathcal{H} \cup R$
    \STATE $S \gets \emptyset$ \COMMENT{Temporary set for the next batch}

    \FORALL{tasks $v \in R$}
        \FORALL{out-neighbors $u$ such that $(v,u) \in E$}
            \STATE $\text{in-degree}(u) \gets \text{in-degree}(u) - 1$
            \IF{$\text{in-degree}(u) = 0$}
                \STATE $S \gets S \cup \{u\}$
            \ENDIF
        \ENDFOR
    \ENDFOR
    \STATE $R \gets S$ \COMMENT{Update ready queue with new available tasks}
\ENDWHILE

\RETURN $\mathcal{H}$

\end{algorithmic}
\end{algorithm}

\subsection{ Fault Isolation Mechanism}
\label{app:faulty}
Runtime failures are isolated to prevent cascading workflow termination.

\begin{algorithm}[H]
\small
\caption{Fault-Isolated Task Execution}
\begin{algorithmic}[1]

\REQUIRE Task $t$
\ENSURE Execution status

\TRY
    \STATE Execute $t$
\CATCH{Exception $e$}
    \STATE Mark $t$ as failed
    \STATE Skip dependent tasks in DAG
\STATE \textbf{end try}

\STATE \RETURN status

\end{algorithmic}
\end{algorithm}


\subsection{ Context Pruning Strategy}
\label{app:context}
To reduce token usage and inference latency, DynAMO selectively propagates context.

\begin{algorithm}[H]
\small
\caption{Keyword-Based Context Selection}
\begin{algorithmic}[1]

\REQUIRE Memory $\mathcal{M}$, keywords $K$
\ENSURE Context $C$

\STATE $C \gets \emptyset$

\FORALL{entries $m \in \mathcal{M}$}
    \IF{$|K \cap m| > 0$}
        \STATE $C \gets C \cup \{m\}$ \COMMENT{Add entry to context}
    \ENDIF
\ENDFOR

\RETURN $C$

\end{algorithmic}
\end{algorithm}


\subsection{ Dataset Details}

Experiments were conducted on \textbf{141 industrial queries} from AssetOpsBench, covering:

\begin{itemize}
    \item IoT sensor retrieval
    \item Failure-mode mapping
    \item Time-series anomaly detection
    \item Work-order generation
\end{itemize}

Concurrency experiments used the full query set.

Context-size experiments used a representative subset of vibration-diagnostics tasks.


\subsection{ Model Configuration}

All planning and reasoning tasks used:

\begin{itemize}
    \item Model: \texttt{meta-llama/\allowbreak Llama-4-Maverick-\allowbreak 17B-128E-Instruct-FP8} (an instruction-tuned mixture-of-experts model with 17B active parameters, served in FP8).
    \item Temperature: 0.0 (deterministic planning)
    \item Max retries: 5
    \item Max workflow steps: 15
\end{itemize}

Latency was measured via wall-clock time.

Results are reported as the median over three runs.


\subsection{ Reproducibility Protocol}

To evaluate stability:

\begin{itemize}
    \item Sequential and parallel workflows were executed 5 times.
    \item Median, standard deviation, minimum, and maximum latency are reported.
    \item Parallel execution reduced variance by over 60\%.
\end{itemize}

All experiments were executed under identical hardware and API configurations to ensure consistency.
\subsection{Sequential and Parallel Workflow Schedulers}
\label{app:schedulers}

DynAMO supports two execution strategies over a planner-generated directed acyclic graph (DAG): 
\textbf{SequentialWorkflow} and \textbf{ParallelWorkflow}. Both operate on the same validated task graph but differ in scheduling policy.

\paragraph{SequentialWorkflow (Strict Topological Execution)-}

SequentialWorkflow executes tasks in a fixed topological order. 
Given a DAG $G=(V,E)$:

\begin{enumerate}
\item Compute a topological ordering of $V$.
\item For each task $v \in V$:
    \begin{itemize}
        \item Construct context from resolved dependencies.
        \item Execute assigned agent(s).
        \item Store output in workflow memory.
    \end{itemize}
\end{enumerate}

This design guarantees deterministic execution but serializes independent tasks, potentially increasing wall-clock latency.

\paragraph{ParallelWorkflow (Dependency-Aware Concurrent Execution)-}

ParallelWorkflow dynamically schedules tasks whose dependencies are satisfied.
Let $\text{deg}^{-}(v)$ denote unresolved dependencies of task $v$.

\begin{enumerate}
\item Initialize ready set $\mathcal{R} = \{v \mid \text{deg}^{-}(v)=0\}$.
\item While $\mathcal{R}$ is non-empty:
    \begin{itemize}
        \item Dispatch all $v \in \mathcal{R}$ concurrently.
        \item Upon completion of $v$:
        \begin{itemize}
            \item Store output in memory.
            \item Update $\text{deg}^{-}$ for dependent nodes.
            \item Add newly satisfied nodes to $\mathcal{R}$.
        \end{itemize}
    \end{itemize}
\end{enumerate}

\paragraph{Execution Properties-}

\begin{itemize}
\item \textbf{Correctness:} Dependency constraints are strictly enforced.
\item \textbf{Determinism:} Sequential execution is fully deterministic; parallel execution is dependency-deterministic.
\item \textbf{Complexity:} Both operate in $O(|V|+|E|)$ time; parallel scheduling reduces wall-clock latency when independent branches exist.
\end{itemize}

\paragraph{Example-}
For tasks $\{S1, S2, S3, S4\}$ with edges 
$S2 \rightarrow S1$ and $S4 \rightarrow \{S2, S3\}$:

Sequential execution:
$S1 \rightarrow S2 \rightarrow S3 \rightarrow S4$.

Parallel execution:
$S1$, then $S2$ and $S3$ concurrently, followed by $S4$.

\end{document}